\documentclass{optica-article}

\journal{opticajournal} % for journals or Optica Open

\articletype{Research Article}

\usepackage{lineno}
\usepackage{physics}
\usepackage{dsfont}

\begin{document}

\title{Quantum Key Distribution With an Integrated Photonic Receiver}

\author{Giulia Guarda,\authormark{1,2} Domenico Ribezzo,\authormark{2,3} Tommaso Occhipinti, \authormark{4} Alessandro Zavatta, \authormark{2,4} and Davide Bacco \authormark{4, 5, *} }

\address{\authormark{1}European Laboratory for Non-Linear Spectroscopy (LENS), 50019 Sesto Fiorentino, Italy\\
\authormark{2}Istituto Nazionale di Ottica, Consiglio Nazionale delle Ricerche (CNR-INO), 50125 Firenze, Italy\\
\authormark{3}University of Naples Federico II, Department of Physics, 80125 Napoli, Italy\\
\authormark{4}QTI S.r.l.,  50125 Firenze, Italy\\
\authormark{5}University of Florence, Department of Physics and Astronomy, 50019 Sesto Fiorentino, Italy}

\email{\authormark{*}davide.bacco@unifi.it} 

% use {asbstract*} to suppress the copyright line. Copyright information will be added in production

\begin{abstract*} 
Photonic integrated circuits (PICs) are key in advancing quantum technologies for secure communications. They offer inherent stability, low losses and compactness compared to standard fiber-based and free-space systems.

Our reasearch demonstrates PIC's effectivness in enhancing quantum communications, implementing a three-state BB84 protocol with decoy-state method. We employ an integrated receiver and superconducting nanowire single photon detectors (SNSPDs) to achieve technological advancements.

One of the most notable results is the extraction of a secret key over a record-breaking 45 dB channel attenuation. Our results demonstrate a remarkable 220\% boost in key rate compared to our prototype fiber-based receiver over a 10 dB channel attenuation. This improvement in the secret key rate (SKR) signifies the potential of integrated photonics to advance the field of quantum communication.
\end{abstract*}

\section{Introduction}
Our society heavily relies on secure communication links for the exchange of information. However, recent advancements in quantum computing pose a significant threat to the security of our well-established network infrastructure \cite{wehner2018quantum}. Consequently, there is a pressing need to develop innovative cryptographic techniques, robust against quantum computational advances. One prominent example of this is Quantum Key Distribution (QKD) \cite{gisin2002quantum},  where the security of the protocols is grounded in the laws of physics and not related to the limited computational power of an eavesdropper \cite{scarani2009security}.  QKD enables the secure distribution of a random secret key among two or more users by harnessing the unique properties of quantum physics \cite{pirandola2020advances}. 
The most studied and widely used protocol to perform QKD is the BB84 protocol, where a prepare and measure scheme ensures the exchange of a secret key exploiting different degrees of freedom of photons, e.g., polarization \cite{liu2010decoy}, time of arrival and phase, \cite{ribezzo2022deploying,ribezzo2023quantum,bacco2019field}, path \cite{da2021path}, etc. 

In this work, we implement an efficient BB84 protocol with time-bin and phase encoding. In this scheme, an imbalanced Mach-Zender Interferometer (iMZI) is required to decode the phase-encoded states. 
We examine the advantages of an integrated iMZI at the receiver's end. Compared to standard free-space \cite{liao2018satellite} and fiber-based \cite{ribezzo2022deploying, ribezzo2023quantum, bacco2019field, guarda} interferometers, our integrated solution is more compact, requires no active stabilization, and is more efficient. Free-space devices are bulky and challenging to maintain stable over extended periods of time, while fiber-based interferometers require additional feedback loops to mitigate drifts in the system, often introducing noise in the detection system and making more complex the hardware implementation. In contrast, the photonic interferometer overcomes these limitations, offering a compact device with low losses and long-term stability \cite{elshaari2020hybrid, wang2020integrated, paraiso2021photonic}.

Integrated platforms represent a practical and scalable choice for increasing the hardware's complexity \cite{adcock2020advances} and pave the way towards a more mature deployment of quantum technologies.
By confining the propagation modes within a solid-state device, PICs significantly reduce noise caused by vibrations and minimize temperature fluctuations \cite{paraiso2021photonic, luo2023recent, cocchi}, making the integrated devices a more stable solution compared to fiber-based systems \cite{silverstone2015qubit, adcock2019programmable}. Moreover, the lithographic fabrication process can be precisely replicated, facilitating scalable mass production of reproducible devices \cite{luo2023recent}.

Currently, state-of-the-art integrated systems are limited by losses and form factors \cite{sax2022high}. Although stability performances are enhanced when using a PIC, there is room for improvements. In this work, we analyze the performances of a PIC-based iMZI at the receiver’s end of a QKD scheme. Hence, we focus on the stability of the interferometer and the propagation losses of the entire integrated device, testing it for the decoding of quantum states in an efficient three-state BB84 protocol.

\section{Methods}
The best-known example of quantum communication is QKD, which enables the distribution of a secret random key between two or more users by exploiting quantum states made by photons. In our work, we implement an efficient three-state BB84 protocol using time-bin encoding and one-decoy state method \cite{ribezzo2022deploying, ribezzo2023quantum, bacco2019field, rusca2018security, boaron2018secure}. In this protocol, the transmitter randomly generates quantum states encoded in two bases, with each quantum state consisting of two-time bins: early and late. 
In contrast to the standard BB84 protocol where both states in the X-basis - $\ket{0}_x$ and $\ket{1}_x$ - are generated, in our scheme only the state $\ket{0}_x$ is produced. Importantly, it has been proved that this improvement does not compromise the security of the protocol \cite{rusca2018finite, hayashi2014security}. However, X-basis states can exclusively be used for security checks and not for generating the final key. This modification brings advantages in terms of reduced complexity in the state generation system and a decreased need for detectors in the decoding system. Moreover, by optimizing the Z-basis choice beyond the balanced probability $p_Z=50\%$, Alice and Bob can reduce the amount of bits to be discarded during the key sifting stage.
The three distinct states can be distinguished by considering both the temporal component and the intensity of the pulses. Mathematically, they are represented as:
\begin{equation}
    \ket{0}\textsubscript{z} = \ket{\gamma}\textsubscript{early} \ket{0}\textsubscript{late};\ \
    \ket{1}\textsubscript{z} = \ket{0}\textsubscript{early} \ket{\gamma}\textsubscript{late};\ \
    \ket{0}\textsubscript{x} = \frac{\ket{0}\textsubscript{z} + \ket{1}\textsubscript{z}}{\sqrt{2}}\ .
\end{equation}

The pulses encoding the states are generated by carving and attenuating a continuous wave C-band laser down to the single-photon level. Therefore, they are weak coherent states and not ideal single photon Fock states. Weak coherent pulses are advantageous for QKD purposes since they can be generated at higher rates than any existing single photon source \cite{grunenfelder2020performance}. However, although they are strongly attenuated, the probability of getting a multi-photon event is not zero, which makes the protocol vulnerable to photon number splitting (PNS) attacks \cite{scarani2009security}.
To address this issue, a technique known as decoy-state method is implemented \cite{hwang2003quantum,lo2005decoy,zhao2006experimental}, where the intensity of the pulses is randomly varied at the transmitter's end.
By doing so, if a PNS occurs, it becomes possible to detect it by looking at the losses of the different intensities pulses, and the quantum security of the protocol is preserved \cite{scarani2009security, pirandola2020advances}.

\begin{figure}[htbp]
\centering\includegraphics[width=13cm]{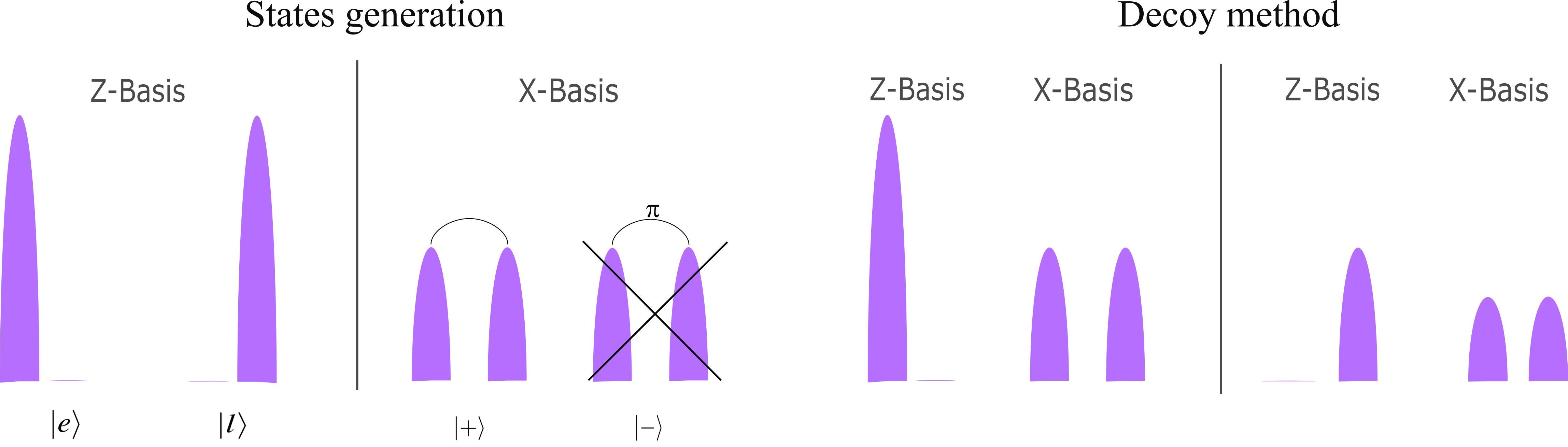}
\caption{\textbf{States generation}. A quantum state exhibits a temporal duration of 1.68 ns and is composed of two time-bins, each separated by a 800 ps time delay. The interval between two consecutive time bins of two consecutive quantum states is 880 ps. The wave function of each photon consists of one or two pulses occurring at the center of either one or both of the time-bins. In the Z-basis encoding, the photons are confined to a single time-bin, while the X-basis states represent the superposition of the Z-basis states. This results in the photon being spread across both time-bins, with a relative phase of either zero or $\pi$.
The established protocol permits the generation of only the phase 0 state. \textbf{Decoy method} A three-level signal sets the different intensities for the decoy.}
\label{fig:states}
\end{figure}

The estimation of the final secret key is done in the finite block-size regime. Here, the upper bound for the secret key rate length ($l$)\ is bounded to \cite{boaron2018secure}:
\begin{equation}\label{eq:skr}
l\leq s_{Z,0}^l+s_{Z,1}^l(1-H_2(\phi_Z^u))-\lambda_{EC}-\lambda_{sec}-\lambda_{corr}.%6\log_2\left(\frac{19}{\epsilon_{sec}}\right)-\log_2\left(\frac{2}{\epsilon _{corr}}\right),
\end{equation}
In Eq.(\ref{eq:skr}), $s_{Z,0}^l$\ and $s_{Z,1}^l$\ are the lower bounds for vacuum and single photon events.
$H_2(x)$\ is the binary entropy, defined as $H_2(x) = -x \ log_2(x) - (1-x) \ log_2(1-x)$.\ \ $\phi_Z^u$\ is the upper bound for the phase error rate, estimated based on the interferometer visibility as $vis_X = 1- 2 QBER_X$.\ As only one state is generated in the X-basis, $\phi_Z^u$\ can only be estimated and not directly measured. Furthermore, $\lambda_{EC}$\ is the amount of disclosed bits during the error correction stage. $\lambda_{sec}$\ and $\lambda_{corr}$ are functions of the security  ($\epsilon_{sec} = 10^{-12}$)\ and correctness ($\epsilon_{corr} = 10^{-12}$)\ parameters in the form of: $\lambda_{sec} = 6\log_2(19 / \epsilon_{sec})$\ and $\lambda_{corr}=\log_2(2 / \epsilon _{corr})$.\ The first quantifies as $\epsilon_{sec} = 10^{-12},$ the probability of the correlation between Eve's and Alice's keys being stronger than the correlation between Alice's and Bob's keys. The second assesses the probability of Alice's and Bob's sifted keys ($S_A$\ and $S_B$)\ differing \cite{canale2014classical}. These can be defined as:
\begin{equation}
    \mathds{1}(S_A,S_B;Z,C) < \epsilon_{sec}; \ \
    P[S_A\neq S_B] < \epsilon_{corr}.
\end{equation}
Here $\mathds{1}(\cdot)$\ is an information measure, $P[\cdot]$\ is a probability function, $Z$\ is the eavesdropper's bit sequence and $C$\ represents the exchanged information.

\section{Experimental setup}
\begin{figure}[htbp]
\centering\includegraphics[width=13cm]{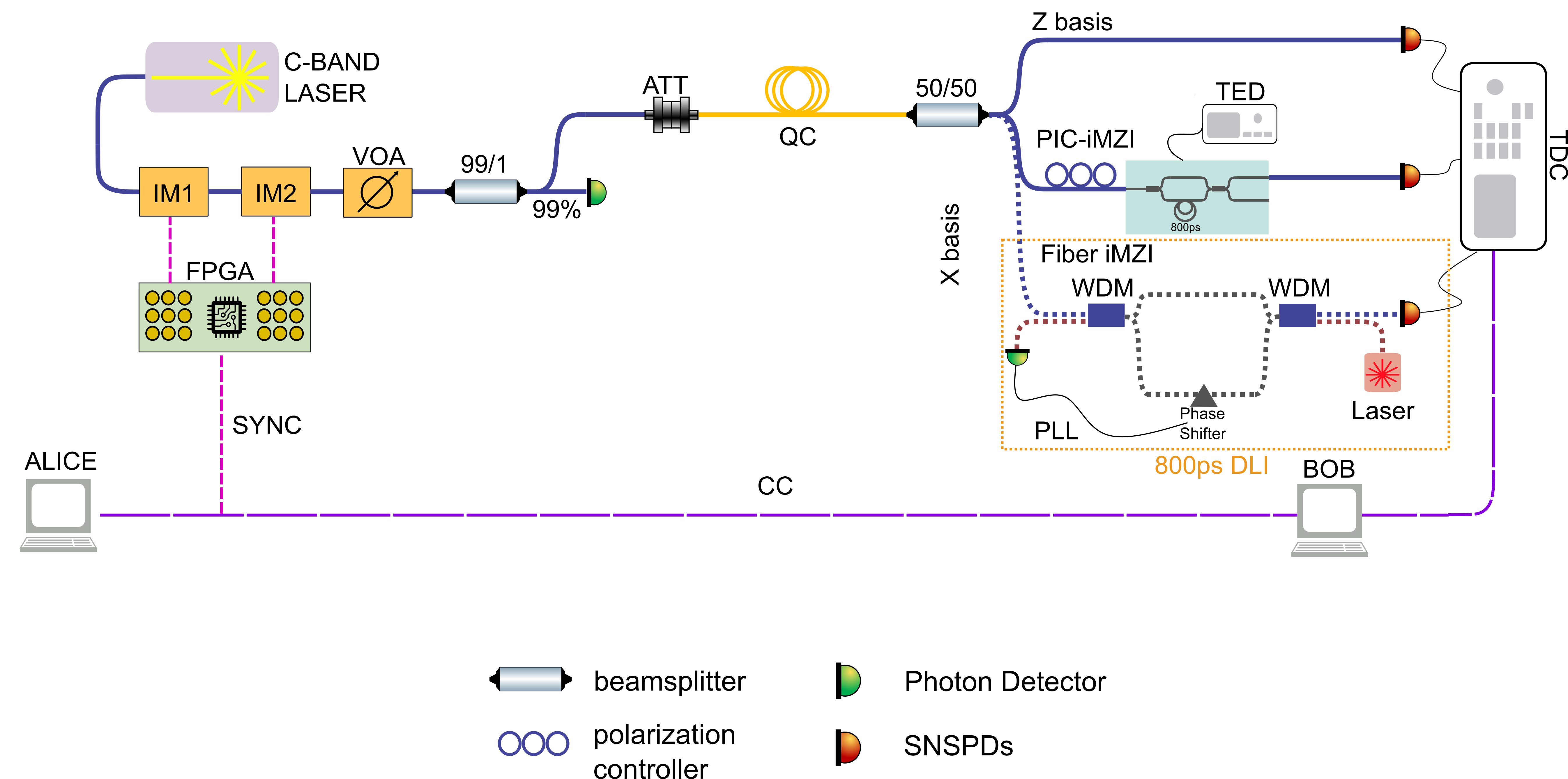}
\caption{\textbf{Setup}. \textbf{ALICE}. A continuous wave laser passes through two intensity modulators (IMs). A field programmable gate array (FPGA) drives the IMs. IM1: states generation, IM2: decoy method. A variable optical attenuator (VOA) is used to set different intensities. A beam splitter adds additional attenuation and enables to monitor the intensities with a photon detector.  An additional fixed attenuation stage (ATT) brings the intensities down to the single photon level. The quantum channel (QC) is made of a single-mode optical fiber and different channel lengths are simulated using fixed optical attenuators. The classical channel (CC) is used for sharing a synchronization signal (SYNC). \textbf{BOB}. A 50/50 beam splitter makes the random detection basis choice. The states measured in the Z-basis go straight to the SNSPD. For the phase measurement in the X basis, the photons need to pass through an imbalanced Mach Zehnder interferometer before being detected. The performances of two different interferometers are compared. \textbf{PIC-based iMZI}. Polarization controllers adjust the polarization of the input light on the PIC, whose performances are polarization-dependent. The fibers are edge coupled. The PIC temperature is stabilized with a thermoelectric temperature controller (TED). \textbf{Prototype fiber-based iMZI} The 800 ps delay line interferometer (DLI) is obtained with careful fiber splicing. A phase-lock loop (PLL) controls the fluctuations and locks the interferometer on a desired phase difference by monitoring the drifts of a counter propagating laser, slightly detuned from the quantum channel wavelength. Finally, the PLL implements a proportional–integral–derivative (PID) controller that acts on a phase shifter in one of the interferometer arms.
SNSPDs used for detection are connected to a time-to-digital converter (TDC), whose data are sent to Bob's computer.Finally, Alice and Bob communicate over the CC for the final generation of a shared secret key.}
\label{fig:setup}
\end{figure}

Figure \ref{fig:setup} shows the architecture of the setup used in this work. In this configuration, the transmitter (Alice) is embedded in a compact 2U rack box. The receiver (Bob) includes a beam splitter for random detection basis choice, and two different interferometers. The first is an integrated interferometer enclosed in a copper case, and temperature stabilized with a thermoelectric temperature controller. The second is a prototype fiber-based interferometer, specifically realized for this experiment. The two parties communicate both via a quantum channel (QC) and a classical channel (CC). The QC is composed by a single-mode fiber at telecom wavelength (SMF-28, ITU G-652) and is used for the transmission of the quantum states. The CC, made of electrical cables and it is employed as the service channel, serving for clock synchronization.
The Bob's detection system includes high-efficient superconducting nanowire single-photon detectors (SNSPDs), with 93\% efficiency, 400 Hz dark count rate and 40 ps of jitter.

On the transmitter side, the states are generated at a repetition rate of 595 MHz by modulating a continuous-wave laser at 1545.32 nm wavelength. In a real-world scenario, the quantum states need to be generated with different random phases to avoid security issues \cite{gottesman2004security,tang2013source}. It does not happen in our transmitter, however, with a view to future field trials, our setup is capable of implementing phase randomization for each state  \cite{francesconi2023efficient}. We employ a field programmable gate array (FPGA) to generate the electrical sequences encoding the states. These signals drive two intensity modulators. One IM shapes the laser signal to generate the time-bin encoding, and the second modifies the states intensities to create the decoy. 
The FPGA generates a sequence of 8190 pulses, corresponding to 4095 states, repeated 145358 times per second. The states are generated in both Z and X bases, with probabilities $p_{Z}^A = 0.5$ and $p_{X}^A = (1-p_Z^A) = 0.5$, respectively. After the generation,  these time-bin-encoded quantum states are attenuated down to the single photon level with a variable optical attenuator (VOA) and additional fixed optical attenuation. Finally, they are transmitted to Bob through the QC.
Different channel lengths are simulated with several fixed optical attenuators.

The decoding scheme at Bob's end involves the use of a 50/50 fiber-based beam splitter for random detection basis selection. Here, probabilities $p_{Z}^B = 0.5$ and $p_{X}^B (1-p_Z^B) = 0.5$ are configured. When the quantum states are measured in the Z-basis, they are directed straight to a SNSPD channel, and the arrival time of the photons is recorded.
In the case of X-basis measurement, an imbalanced Mach Zehnder interferometer (iMZI) is necessary to properly measure the phase. In this work, we compare the performances of a lab-made fiber-based iMZI with a PIC-based one. The fiber-based iMZI has been constructed specifically to conduct this experiment. It requires active stabilization, done with a phase lock loop (PLL). This consists of a monitor signal at 1550.00 nm, entering from the iMZ output and counter-propagating with respect to the quantum states path, a piezo-controlled phase shifter, and an InGaAs photodetector. A proportional–integral–derivative (PID) controller, acting on the phase shifter, is in charge for keeping the phase difference of the two interferometer arms constant. As the PLL wavelength is just slightly different from the quantum states one, locking the phase of the monitoring signal enables concurrent phase locking for the quantum signal. Both the input and output port of the interferometer are equipped with a wavelength division multiplexer (WDM) in charge for combining and dividing the two signals. The hardware is supported by an auto-tuning software that hinders drifts in the setup.
The PIC-based iMZI, on the other hand, does not require  an active stabilization system. The quantum signal is coupled into the PIC via edge coupling method. As the waveguides are polarization-dependent, polarization controllers are used to control the incoming photons polarization. Once coupled into the waveguides, the two pulses composing a time-bin are overlapped using an 800 ps delay line into one of the arms of the interferometer. The phase between the two arms is adjusted and kept stable using a thermoelectric cooler (TEC) controlled by a temperature controller (TED) using a PID loop.  Subsequently, the signal is collected into an optical fiber, edge-coupled, and glued to the PIC before being directed to the SNSPD. 

To ensure accurate data collection, the signals are recorded using a time-to-digital converter (TDC), a qutag time-tagging unit produced by qutools GmBH, and sent to Bob's computer. Finally, Alice and Bob communicate using the CC to obtain a quantum criptographically secure secret key. The block size, $n_Z$,\ which determines the number of bits used for the finate key analysis, is set to $10^7$.\ The data are collected for a full block time, i.e. the time required to acquire a full block size, $\tau_Z$.\ The longest acquisition time was done for the longest simulated channel, 45 dB. At this attenuation level, it was required a block time of 3 hours and 15 minutes of continuous acquisitions.

\subsection{Integrated interferometer}
The PCI iMZI utilized in this work is made of passive components. The platform is composed of a borosilicate glass matrix, specifically designed for making ion exchange waveguides, with an effective refractive index of 1.5212 at 1545.32 nm \cite{broquin2021integrated}.

By inducing a localized alteration in the glass matrix composition, it is possible to produce changes in the refractive index, allowing for the creation of waveguides. This process involves thin-film deposition, photolithography, and a chemical ion-exchange bath. The resulting waveguides exhibit minimal losses and seamless compatibility with conventional optical fibers, making them well-suited for telecommunication applications \cite{broquin2021integrated}.

Once the temperature of the PIC is stabilized, using a PID loop, it remains constant for the whole measurement time, as it is enclosed in a copper case ensuring thermal stability. This long-term stability enables us to conduct continuous measurements lasting for 2 days.

\section{Results and Discussion}
We compare the stability and secret key rate of two receivers. The first receiver uses a fiber-based interferometer to measure the X-basis \cite{guarda}. In contrast, the second receiver employs a fully photonic integrated interferometer. The advantages of the integrated solution are evident in both stability comparisons and amount of secret key extracted from the two setups under various attenuation channels.

\begin{figure}[htbp]
\centering\includegraphics[width=13cm]{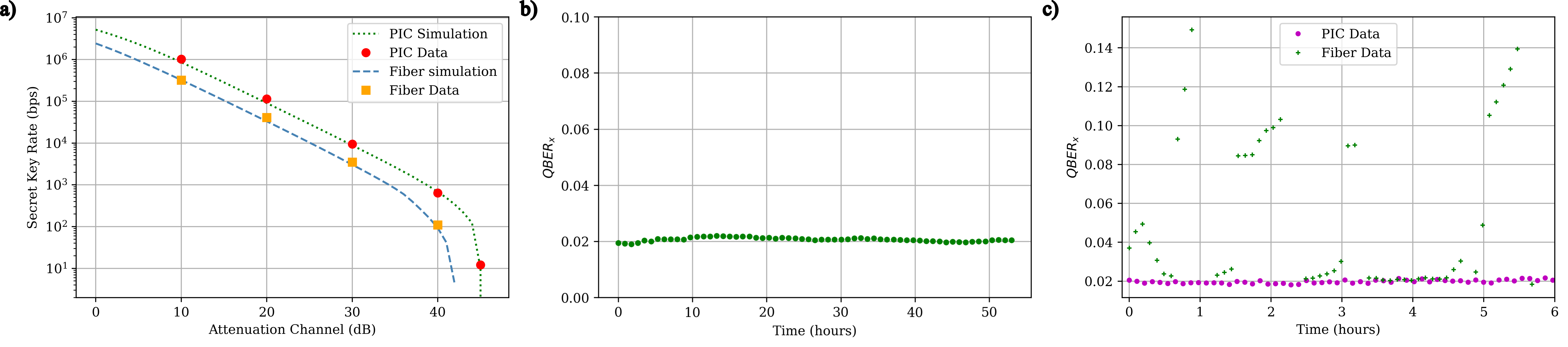}
\caption{\textbf{a) Secret key rate as a function of the channel losses}. The dotted green line represents the extracted key rate for different channel attenuations for the PIC-based interferometer according to our simulation model. The red circles are the measured SKR values for the same system. The dashed blue curve is the simulation for the fully fiber-based interferometer and the orange squares are the measured values for the same system. \textbf{b) Stability}. Trend of \textit{QBER\textsubscript{x}} during the key exchange over a channel showing 30 dB of attenuation for over 50 hours of continuous measurements. The green circles are the data relative to the PIC-based iMZI. \textit{QBER\textsubscript{x}} is connected to the interferometer visibility \textit{vis} by $QBER_X=(1-vis)/2$. \textbf{c) Stability comparison}. The trend of the QBER\textsubscript{x} during the key exchange at 30 dB for both the fiber-based and the PIC-based interferometer. The purple circles are the date relative to the PIC-based iMZI and the green cross shaped points are relative to the fully fiber-based iMZI. All the experimental points have been acquired for a full block time.}
\label{fig:results}
\end{figure}

The advantages of the PIC stem from its ultra-low losses and inherent stability. On the other hand, the fiber-based interferometer relies on a PLL to maintain stability \cite{ribezzo2023quantum}. This mitigates the interferometer fluctuations but introduces noise due to the counter-propagating monitor signal, which hinders secret key extraction at longer distances. Furthermore, the enhanced integrated device stability and noise reduction, lead to increased key extraction rates for different channel lengths, as illustrated in Fig.\ref{fig:results}\textbf{a)}.

The PIC-based interferometer demonstrates superior stability. Even when the fiber-based setup achieves good stability, the PIC-based interferometer registers lower Quantum Bit Error Rate (QBER) values, due to its higher extinction ratio.
Fig.\ref{fig:results}\textbf{b)} illustrates the long-term stability achieved with the integrated iMZI, demonstrating over 50 hours of continuous measurements. This extended stability allows for extended data acquisition over a period of more than two days.

Fig.\ref{fig:results}\textbf{c)} illustrates the stability of the fiber-based iMZI during a 5-hour continuous data acquisition at 30 dB attenuation. As the system drifts, the PLL requires an internal re-calibration, which takes some time, leading to uneven time intervals between data points. A comparison with the integrated interferometer highlights the significant benefits of long-term stability.

Different materials for photonic integration have been studied. The most common include silicon-based platforms \cite{lockwood2013light} such as silicon and silicon nitride \cite{sibson2017integrated,sibson2017chip}; nonlinear materials such as lithium niobate \cite{domeneguetti2023fully}; and glass waveguides, such as aluminium borosilicate glass \cite{sax2022high}.
In a standard quantum communication protocol, quantum states undergo manipulation through a combination of active and passive components.
Common discrete variables encoding methods include polarization, path, and time-bin and phase degrees of freedom \cite{luo2023recent}, and they generally exploit active components. When it comes to decoding the quantum information, passive components come into play. Among these, a widely used configuration involves interferometers, which incorporates elements like phase shifters, delay lines, and beam splitters \cite{orieux2016recent}. 

Tab.\ref{tab:table1} reports the most impactful solution adopted for a BB84 time-bin receiver. Different loss features, material choices, and form factors (FFactors) are reported.
\begin{table}[h]
   \centering
\caption{\textbf{State-of-the-art}.\\
\textsuperscript{($\ast$)}: aluminum borosilicate glass, \textsuperscript{($\dagger$)}: borosilicate glass.} \label{tab:table1}
\begin{tabular}{c|c|c|c}
         
         Ref & Loss & FFactor & Material\\
         \hline \cite{sibson2017integrated}  & 4.5 dB  & 2.5 mm & SiO\textsubscript{x}N\textsubscript{y} \\
         \hline  \cite{sibson2017chip} & 9 dB & 32 mm & SiO\textsubscript{x}N\textsubscript{y} \\
         \hline  \cite{sax2022high} & 3.50 dB & 150 mm & \textsuperscript{($\ast$)} \\
         \hline Ours  & 2.75 dB & 10 mm & \textsuperscript{($\dagger$)} \\
         
\end{tabular}
\end{table}%

Finally, Tab.\ref{tab:table2} shows a comparison with state-of-the-art devices.
\begin{table}[h]
   \centering
\caption{\textbf{Results}.} \label{tab:table2}
\begin{tabular}{c|c|c|c}
         
         Ref & Length & Att & SKR  \\
         \hline \cite{sibson2017integrated}  & 20.0 km & & 916.0 kbps \\
         \hline \cite{sibson2017chip} & 20.0 km & & 568.0 kbps \\
         \hline \cite{sax2022high} & & 39.5 dB & 940.0 bps \\
         \hline Ours  & & 45.0 dB & 12.2 bps  \\
         
\end{tabular}
\end{table}%

Overall it is possible to extract a secret key from the fully prototype fiber-based system up to 40 dB of attenuation channel, equivalent to 205 km in low-loss fibers, at a rate of 110 bps. On the other hand, the chip-based device extracts a positive key at a rate of 637.8 bps at 40 dB of attenuation channel, and 12 bps with 45 dB attenuation. This is equivalent to a distance of 225 km in standard low-loss fibers and represents, to the best of our knowledge \cite{sax2022high, liu2022advances}, the longest distance covered with a fully-integrated receiver system.

\section{Conclusion}
In summary, our work showcases the pivotal role of photonic integrated circuits in propelling the boundaries of quantum communication. 

The system, supported with a PIC-based iMZI, shows a QBER as low as 2\%. This value is below what we have previously observed for standard fully fiber-based receivers \cite{ribezzo2022deploying, ribezzo2023quantum, guarda,bacco2019field}, thanks to low losses in the setup and higher stability in the interferometer. These features (stability and low-losses) allow for over 2 days of continuous measurements with excellent stability. Finally, the integrated receiver enables the extraction of a secret key up to 45 dB, equivalent to a distance of 225.0 km in standard low-loss optical fibers, the furthest distance achieved nowadays with a fully integrated device \cite{sax2022high, liu2022advances}.

\begin{backmatter}
\bmsection{Acknowledgments}
This work was funded by the European Union (ERC, QOMUNE, 101077917, by the Project EQUO (European QUantum ecOsystems) which is funded by the European Commission in the Digital Europe Programme under the grant agreement No 101091561, the Project SERICS (PE00000014) under the MUR National Recovery and Resilience Plan funded by the European Union - NextGenerationEU, the Project QuONTENT under the it Progetti di Ricerca, CNR program funded by the Consiglio Nazionale delle Ricerche (CNR) and by the European Union - PON Ricerca e Innovazione 2014-2020 FESR - Project ARS01/00734 QUANCOM, the Project QUID (Quantum Italy Deployment) funded by the European Commission in the Digital Europe Programme under the grant agreement No 101091408.
\end{backmatter}

%%%%%%%%%%%%%%%%%%%%%%% References %%%%%%%%%%%%%%%%%%%%%%%%%

\bibliography{bibliography}

\end{document}